\documentclass[preprint,nofootinbib,aps,superscriptaddress,eqsecnum]{revtex4-1} %\documentclass[11pt, a4paper]{article}
\usepackage{float}
\pdfoutput=1
\usepackage{extarrows}
\usepackage{graphicx}
\usepackage{amsmath}
\usepackage{amsfonts}
\usepackage{amssymb}
\usepackage{hyperref}
\usepackage{caption}
\usepackage{subcaption}
\usepackage{color}
\allowdisplaybreaks
\def\bea{\begin{eqnarray}}
\def\eea{\end{eqnarray}}
\def\be{\begin{equation}}
\def\ee{\end{equation}}

\begin{document}

	\title{Phenomenology of one zero texture Yukawa matrix in a flavor symmetric scotogenic model}
	
		\author{Lavina Sarma}
	\email{lavina@tezu.ernet.in}
\author{Mrinal Kumar Das}
	\email{mkdas@tezu.ernet.in}
	\affiliation{Department of Physics, Tezpur University, Tezpur 784028, India}

	\begin{abstract}
	The scotogenic model is well known for accommodating small neutrino mass and dark matter. Here, we have realized the scotogenic model with the help of discrete flavor symmetry $A_{4}\otimes Z_{4}$. We have obtained three one zero textures of Yukawa coupling matrices from the model and have studied its impact on neutrino phenomenology and related aspects of cosmology. On the basis of $\mu-\tau$ symmetry, we have further discarded two structures of one zero texture Yukawa coupling matrix. We further analyze if the effective mass of active neutrinos obtained by the virtue of the Yukawa coupling matrix is consistent with the KamLAND-Zen limit for 0$\nu\beta\beta$. Also different lepton flavor violating(LFV) proceses such as $l_{\alpha}\longrightarrow l_{\beta}\gamma$ and $l_{\alpha}\longrightarrow 3l_{\beta}$ are implemented and their influence on neutrino phenomenology is studied corresponding to the Yukawa coupling matrix. The entire work is carried out considering the dark matter mass($M_{DM}$) in the region $450-750$ GeV. We have also obtained some significant results for baryon asymmetry of the Universe in agreement with the one zero textures of the coupling matrices. Furthermore, interesting results for relic abundance on the basis of distinct mass splittings between the inert scalars. \\

	\end{abstract}
	\maketitle	
	
	\section{Introduction}
	
	We are familiar with the immense accomplishment of the Standard Model(SM) in explaining the theory for fundamental particles and its interactions. Inspite of being an affluent and self-consistent model, it is certainly incomplete. Various observations point towards the need for physics beyond the Standard Model(BSM). This includes the non-zero mass of neutrinos\cite{deSalas:2017kay}, Baryon Asymmetry of the Universe(BAU)\cite{leptogenesis,Hugle:2018qbw,neutrinomasspdg,Minkowski,Mohapatra,Yanagida:1979as,Schechter:1980gr,Glashow:1979nm}, Dark Matter(DM)\cite{bertone2005particle,Moore:1999nt}, etc. The neutrino oscillations\cite{neuOsc,neuOsc2} has revealed that the neutrinos are massive\cite{Lattanzi:2016rre,Senjanovic} and also the fact that their flavors mix. The recent Neutrino experiments MINOS\cite{MINOS},RENO\cite{RENO},T2K\cite{T2K},Double-Chooz\cite{DCHOOZ} have not only confirmed but also measured the neutrino oscillation parameters more accurately\cite{Choubey}. There are many crucial evidences of DM as mentioned in these literatures \cite{Zwicky:1933gu,Treu:2012sn}. The most significant cosmological bound comes from Planck satellite \cite{Ade:2015fva} which suggests that approximately $27\%$ of the present Universe is comprised of DM. Together with the cosmic baryon asymmetry of the Universe, we have explicit reasons to extend the SM with new particles and fields.
	We are well aware of the various beyond the Standard model frameworks which tends to incorporate the explainations for the above mentioned anamolies. This includes the seesaw mechanisms such as type I\cite{Minkowski}, type II\cite{ANTUSCH2004199}, type III\cite{Foot1989}, inverse\cite{HIRSCH2009454,khalil2010tev} and radiative seesaws\cite{Ma11,Ma:2006km,Ma:2017kgb}. \hspace{0.5cm}
	
	In our work, we mainly focus on the radiative seesaw mechanism which is of much significance in connecting neutrino and dark matter phenomenology. We consider the scotogenic model\cite{Scotogenic} which is an extension of the SM by three heavy neutral singlet fermions and an inert scalar doublet. These extra fields in the extension are experimentally observable at the forthcoming Large Hadron Collider(LHC), with an important implication that the lightest of them could be a significant candidate for the dark matter of the Universe. We basically realise the generic scotogenic model with the help of discrete flavor symmetry $A_{4}\otimes Z_{4}$\cite{LS}. With proper choice of vacuum expectation value(vev), allowed by $A_{4}$ symmetry\cite{Zhang}, we are able to generate three structures of Yukawa coupling matrices with one zero texture. We further investigate the phenomenology related with these matrices in both neutrino as well as cosmology sector. A comparative study between the three cases is carried out, thereby determining which one could be viable in satisfying the bounds from various observations. Neutrinoless double beta decay process\cite{Mohapatra:1986su,Barry:2013xxa} is evaluated to check if the effective mass of the active neutrinos abide by the limit given by KamLAND-Zen\cite{Kamland2,kamland}. Consecutively, we have studied the lepton flavor violating(LFV) proceses such as $l_{\alpha}\rightarrow l_{\beta}\gamma$ and $l_{\alpha}\rightarrow 3l_{\beta}$, to examine their impression on the neutrino phenomenology. The most stringent bounds on LFV comes from the MEG experiment\cite{TheMEG} giving limit on Br($\mu \rightarrow e\gamma$)$< 4.2\times 10^{-13}$. In case of $l_{\alpha}\rightarrow 3 l_{\beta}$ decay, bound from SINDRUM experiment\cite{Perrevoort:2018cqi} is set to be $\rm BR(l_{\alpha} \rightarrow 3l_{\beta})<10^{-12}$. We have analysed $N_{1}$ leptogenesis\cite{leptogenesis,Hugle:2018qbw,BBB} in all the three cases of Yukawa coupling matrices as it is has a direct consequence in the generation of baryon asymmetry of the Universe. The mass of the lightest heavy neutral singlet fermion is of TeV scale, which has a lower limit $M_{1} \simeq 10$ TeV\cite{Hugle:2018qbw,Mahanta:2019gfe}. Thus, a low scale leptogenesis, generally termed as vanilla leptogenesis is also possible in a scotogenic model unlike other seesaw mechanisms.
	Our work is primarily carried out with the dark matter candidate(in our case, lightest of the inert scalar doublet) having mass in the regime $450-750$ GeV. However, as seen in various literatures\cite{Ma:2006km,LopezHonorez:2006gr,Ahriche:2017iar}, in the IHDM desert, i.e., $M_{W} < M_{DM} \le 550$ GeV, the generation of relic abundance is prohibited. But, with proper choice of the mass splitting between the scalars and fine tunning of the quartic coupling, it is possible to get the observed relic abundance for $400\leq M_{DM}\leq 550$ GeV. \\
We know that the Yukawa coupling in scotogenic model responsible for generating neutrino mass and freeze-out of dark matter, also significantly persuade lepton flavor violating processes such as $l_{\alpha}\rightarrow l_{\beta}\gamma$ and $\mu \longrightarrow e$ conversion at one loop level. In order to naturally suppress these decays, we can consider the choice of parameters, mainly the elements of Yukawa coupling matrix to play a vital role. One such possibility is obtained by assigning two zeroes simultaneously in the Yukawa coupling matrices. However, this choice leads to a disfavoured range of the UPMNS mixing angles according to the $3\sigma$ global fit data\cite{texture1}. Therefore, we take the assumption of one zero texture in Yukawa coupling matrix to obtain a supressed lepton flavor violating processes simultaneously obeying the $3\sigma$ range for neutrino oscillation parameters. Furthermore, we proceed the entire phenomenological study taking into consideration of the one zero texture Yukawa coupling matrix which is more preferable as also studied in\cite{texture1,texture2}. 
In various neutrino mass models as studied in the literatures\cite{Gautam,Borgohain}, the allowance of two zero texture on basis of neutrino oscillation parameters is obtained. It is seen to abide by the KamLAND-Zen limit for effective neutrino mass as well. Also as mentioned in\cite{Gautam}, where the phenomenological study of texture zeroes in (2,3) inversee seesaw is carried out, we see that the higher texture zero structures are disallowed, whereas all the possible structures of texture two zero is successful in producing the desired results for $0\nu\beta\beta$ and relic abundance of dark matter candidate.

	We have further categorised the paper into seven sections which are as follows. Sec.\eqref{sec2} and sec.\eqref{sec3} includes the generic scotogenic model and the flavor symmetric scotogenic model respectively. We analyse the model by verifying it under the various phenomenological constraints as mentioned in sec.\eqref{sec4} Phenomenologies such as neutrinoless double beta decay, lepton flavor violation, leptogenesis, and dark matter are therefore explicitly mentioned in subsections \eqref{A}, \eqref{B}, \eqref{C} and \eqref{D} respectively. We finally show the results and numerical analysis of our work in sec.\eqref{sec7}, finally followed by the conclusion in sec.\eqref{sec8}.

	\section{Scotogenic model}\label{sec2}
	Scotogenic model is a popular model which can accomodate neutrino mass and dark matter phenomenology simultaneously\cite{LS1,Borah:2018rca}. It is basically an extension of the inert Higgs doulet model (IHDM)\cite{Honorez:2010re,LopezHonorez:2006gr,Arhrib:2012ia,Bhattacharya:2019fgs,BorahBR}. The scotogenic model comprises of an extra scalar field $\eta$ and three neutral singlet fermions $N_{i}$ with $i=1,2,3$ along with the SM particles. $N_i$ and $\eta$ are odd  under a built-in discrete $Z_{2}$ symmetry \cite{Ma:2006km,Hambye:2009pw,Dolle:2009fn,Honorez:2010re,Gustafsson:2012aj}, whereas the SM fields are $Z_{2}$ even. Particles under $Z_{2}$ symmetry transform as follows-
	
	\begin{equation}
	N_{i}\longrightarrow -N_{i},~ \eta\longrightarrow -\eta,~ \phi\longrightarrow \phi,~\Psi\longrightarrow \Psi,
	\end{equation}
	where $\eta$ is the inert Higgs doublet, $\phi$ is the SM Higgs doublet and $\Psi$ denotes the SM fermions.
	Under the group of symmetries $SU(2) \times U(1)_{Y} \times Z_{2}$, the leptonic and scalar particle content can be represented as follows :
	\begin{equation*}
	\begin{pmatrix}
	\nu_{\alpha}\\
	l_{\alpha}
	\end{pmatrix}_{L} \sim (2,  -\dfrac{1}{2}, +) ,~ l^{c}_{\alpha} \sim (1,1,+), ~
	\begin{pmatrix}
	\phi^{+}  \\
	\phi^{0} \\
	\end{pmatrix} \sim (2,\frac{1}{2},+), 
	\end{equation*}
	\begin{equation}
	N_{i} \sim (1,1,-), ~
	\begin{pmatrix}
	\eta^{+}\\
	\eta^{0}\\
	\end{pmatrix} \sim (2,1/2, -).
	\end{equation}
	The scalar doublets are written as follows :
	\begin{equation}
	\eta=
	\begin{pmatrix}
	\eta^{\pm}\\
	\frac{1}{\sqrt{2}}(\eta^{0}_{R} + i\eta^{0}_{I}) \\
	\end{pmatrix}, \quad
	\phi=
	\begin{pmatrix}
	\phi^{+}\\
	\frac{1}{\sqrt{2}}(h+i\xi)
	\end{pmatrix}.
	\end{equation} 
	\hspace{3cm}
	\begin{figure}[h]
		\centering
		\includegraphics[width=0.4\textwidth]{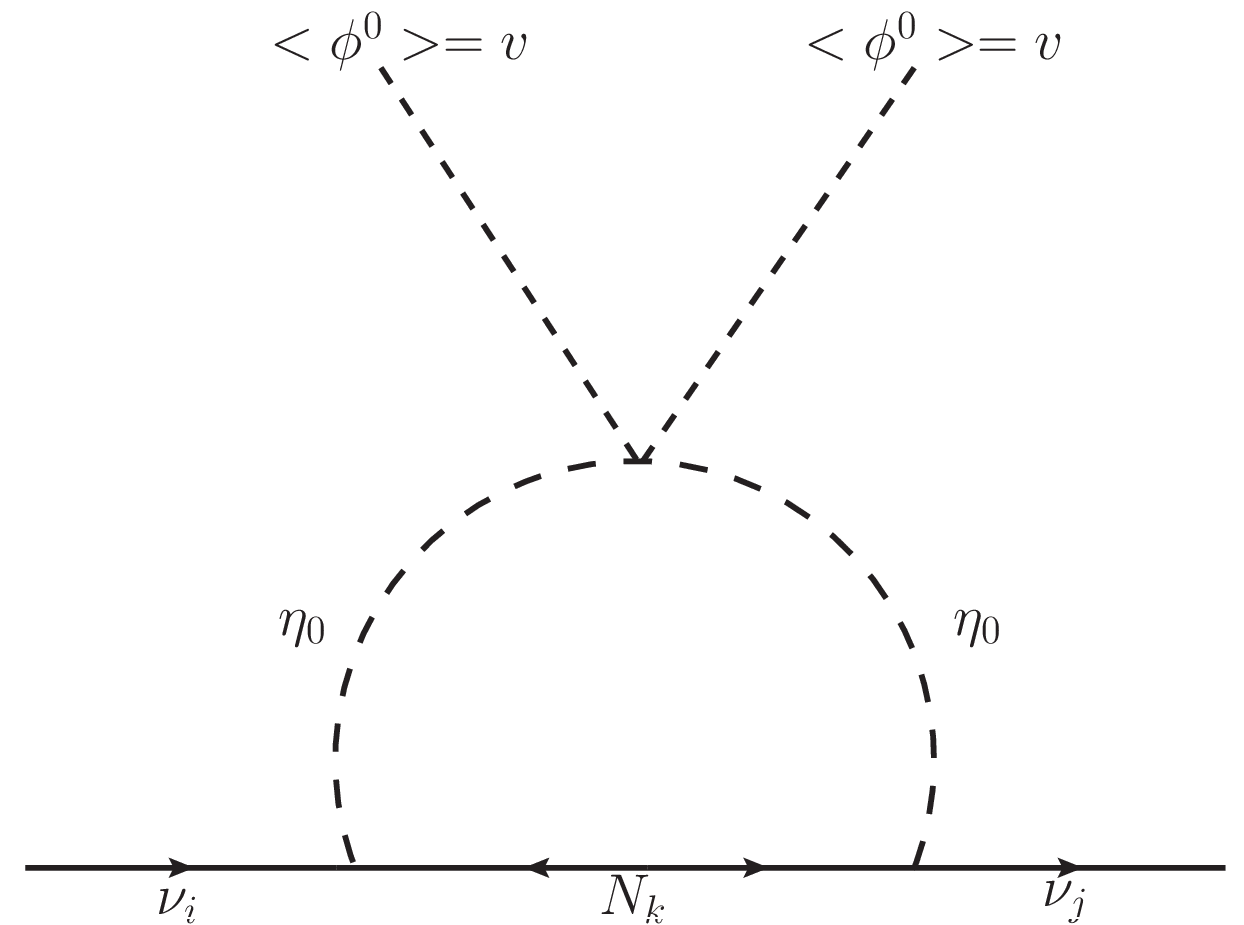}
		\caption{ Mass generation of light neutrino via one-loop contribution by the exchange of right handed neutrino $N_i$ and the scalar $\eta_0$. } \label{fig1}
	\end{figure}\\
As mentioned in the figure\ref{fig1}, we obtain the neutrino mass through a one-loop mechanism. The two Higgs field $\phi^{0}$ in the figure will not propagate but will acquire VEV after the EWSB.
	The lagrangian involving the newly added field is :
	\begin{equation}
	\mathcal{L} \supset \frac{1}{2}(M_{N})_{ij}N_{i}N_{j} + Y_{ij}\bar{L}
	\tilde{\eta}N_{j} + h.c 
	\end{equation}
	The new potential on addition of the new inert scalar doublet is:
	\begin{equation}
	\begin{split}
	V_{Scalar} = &m_{1}^{2}\phi^{+}\phi + m_{2}^{2}\eta^{+}\eta + \frac{1}{2}\lambda_{1}(\phi^{+}\phi)^{2} + \frac{1}{2}\lambda_{2}(\eta^{+}\eta)^{2}+ \lambda_{3}(\phi^{+}\phi)(\eta^{+}\eta) \\ &+
	\lambda_{4}(\phi^{+}\eta)(\eta^{+}\phi) 
	+\frac{1}{2}[\lambda_{5}(\phi^{+}\eta)^{2} + h.c.]\label{eqp1}
	\end{split}
	\end{equation}
	Except for $\lambda_{5}$, Eq. \eqref{eqp1} contains parameters which are real. The masses of the physical scalars: Higgs boson, a pair of charged scalar ($\eta^{\pm}$), CP even scalar($\eta^{0}_{R}$) and CP odd scalar($\eta_{I}^{0}$) are as follows:
	\begin{equation}
	\begin{split}
	m^{2}_h =& -m^{2}_{1} = 2\lambda_{1}\textit{v}^{2},\\
	m^{2}_{\eta^{\pm}} =& m^{2}_{2}+\lambda_{3}\textit{v}^{2},\\
	m^{2}_{\eta_{R}^{0}} =& m^{2}_{2} + (\lambda_{3}+\lambda_{4}+\lambda_{5})\textit{v}^{2},\\
	m^{2}_{\eta_{I}^{0}} = &m^{2}_{2} + (\lambda_{3}+\lambda_{4}-\lambda_{5})\textit{v}^{2}.
	\end{split}
	\end{equation}
 The condition $\lambda_{5} < 0$ is significant as the probable DM candidate is the CP-even scalar. Also, the degeneracy in masses of the neutral components of the inert doublet arises due to the limit $\lambda_{5}\rightarrow 0$\cite{tHooft:1980xss}.

 The neutrino mass matrix arising from the radiative mass model is given by :
	\begin{equation}
	\begin{split}
	\textit{M}_{ij}^{\nu}=&\sum_{k} \frac{Y_{ik}Y_{jk}}{32\pi^{2}}M_{k}\left[\frac{m_{\eta_{R}^{0}}^{2}}{m_{\eta_{R}^{0}}^{2}-M^{2}_{k}}\:ln\frac{m_{\eta_{R}^{0}}^{2}}{M^{2}_{k}}-\frac{m_{\eta_{I}^{0}}^{2}}{m_{\eta_{I}^{0}}^{2}-M^{2}_{k}}\:ln\frac{m_{\eta_{I}^{0}}^{2}}{M^{2}_{k}}\right]\\
	\equiv& \sum_{k} \frac{Y_{ik}Y_{jk}}{32\pi^{2}}M_{k}[L_{k}(m^{2}_{\eta_{R}^{0}}) - L_{k}(m^{2}_{\eta_{I}^{0}})],\label{eq7}
	\end{split}
	\end{equation}
	where $M_{k}$ represents the mass eigenvalue of $N_{k}$(k=1,2,3) with j=1,2,3 representing the three neutrino generation and Y is the Yukawa coupling matrix. The function $L_{k}(m^{2})$ used in Eq. \eqref{eq7} is given by:
	\begin{equation}
	L_{k}(m^{2})= \frac{m^{2}}{m^{2}-M^{2}_{k}}\ln\frac{m^{2}}{M^{2}_{k}}
	\end{equation}

	\section{$A_{4}\otimes Z_{4}$ symmetric scotogenic model}\label{sec3}
	With the help of discrete flavor symmetry, in our case $A_{4}\otimes Z_{4}$, we realise the minimal scotogenic model and obtain the neutrino mass at 1-loop level with the DM candidate contained. The discrete symmetries, i.e $A_{4}\otimes Z_{4}$ imposes significant bounds on the Yukawa coupling matrix which further impacts the model parameters. In our work, we obtain three cases of one zero texture Yukawa coupling matrix by the virtue of the choice of the vev. Furthermore, using these distinct Yukawa coupling matrices, we determine the neutrino mass and analyse the phenomenologies associated with it. The particle content and their respective charges corresponding to the discrete symmetries are given in Table \ref{TAB}. 
	\begin{table}[h]
		
		\begin{center}
			\begin{tabular}{|c|c|c|c|c|c|c|c|c|c|c|c|c|c|c|c|}
				
				\hline 
				Field &$l$ & $e_{R}$ & $\mu_{R}$ & $\tau_{R}$ & $\phi$ &  $\eta$ & $\chi$ & $\chi^{'}$ &$\chi^{''}$ & $\Phi$ & $\kappa$ & $\kappa^{'}$ &  $N_{1}$ & $N_{2}$ & $N_{3}$ \\ 
				\hline
				$(SU(2),U(1)_Y)$ &(2,-1/2)&(1,1)&(1,1)&(1,1)&(2,1/2)&(2,1/2)&(1,0)&(1,0)&(1,0)&(1,0)&(1,0)&(1,0)&(1,1)&(1,1)&(1,1)\\
						\hline
				$A_{4}$& $3$ & $1$ & $1^{''}$ &$1^{'}$& $1$ & $1$ & $3$ &  $3$ &$3$&$3$ & $1$ & $1^{'}$& $1$& $1^{'}$&$1$ \\
				\hline 
				$Z_{4}$& $1$ & $i$ &$i$& $i$ & $1$ & $1$ &  $1$ &$i$&$-1$&$-i$& $1$& $-1$& $1$& $-i$& $-1$\\
				\hline 
			\end{tabular} 
		\end{center}
	\caption{	Fields and their respective transformations
			under the symmetry group of the model.} \label{TAB}
	\end{table}
We have the $A_{4}\otimes Z_{4}$ invariant Lagrangian for the lepton sector as follows:
\begin{equation}
\begin{split}
	\mathcal{L} \supset & \frac{y_{e}}{\Lambda}(\bar{l}\phi\Phi)e_{R}+\frac{y_{\mu}}{\Lambda}(\bar{l}\phi\Phi)\mu_{R}+\frac{y_{\tau}}{\Lambda}(\bar{l}\phi\Phi)\tau_{R}+ \frac{y_{1}}{\Lambda}(\bar{l}\eta\chi)N_{1}+\frac{y_{2}}{\Lambda}(\bar{l}\eta\chi^{'})N_{2}+\frac{y_{3}}{\Lambda}(\bar{l}\eta\chi^{''})N_{3}+ \\ & \frac{1}{2}\omega_{1}\kappa\bar{N_{1}^{c}}N_{1}+\frac{1}{2}\omega_{2}\kappa^{'}\bar{N_{2}^{c}}N_{2}+\frac{1}{2}\omega_{3}\kappa\bar{N_{3}^{c}}N_{3}\label{eq1}
	\end{split}
\end{equation}\label{11}
 With the choice of vacuum expectation value of the flavon $\Phi$, i.e. $<\Phi>=(u,0,0)$\cite{Zhang}, we obtain the flavor structure for charged lepton coupling matrix to be a diagonal one. The charged lepton mass matrix is given by:
\begin{equation}
M_{l}=\frac{<\phi> u}{\Lambda}\left[\begin{array}{ccc}
y_{e}&0&0\\
0&y_{\mu}&0\\
0&0&y_{\tau}
\end{array}\right]\\
\end{equation}
Because of the additional $Z_{4}$ symmetry, the right handed neutrino mass matrix will be diagonal one with $<\kappa>=<\kappa^{'}>=<u>$. 
\begin{equation}
M_{R}=\left[\begin{array}{ccc}
\omega_{1}u&0&0\\
0&\omega_{2}u&0\\
0&0&\omega_{3}u
\end{array}\right]\\,
\end{equation}
where $\omega_{1}$,$\omega_{2}$ and $\omega_{3}$ are the couplings between the right handed neutrinos and the scalar fields $\kappa$ and $\kappa^{'}$(i.e. they are the Yukawa couplings).
As we have mentioned earlier that we tend to obtain three one zero texture Yukawa coupling matrices from this model. This is possible depending on the choice of vev assigned to the flavons $\chi$, $\chi^{'}$ and $\chi^{''}$. The Yukawa coupling matrices that we showcase below are a manifestation of the Dirac mass matrices.\\
Case I: Incorporating the following flavon allignments:
$<\chi>=(0,-u,u)$, $<\chi^{'}>=(u,u,u)$ and $<\chi^{''}>=(u,u,u)$ in Eq.\eqref{eq1}, we obtain a zero term in $Y^{'}_{11}$ position given by:
\begin{equation}
Y^{'} = \left[\begin{array}{ccc}
0 & b & c \\
-a & b & c\\
a & b & c\\
\end{array}\right]\\,\label{Y11}
\end{equation}
where $a=y_{1}\frac{u}{\Lambda}$, $b=y_{2}\frac{u}{\Lambda}$ and $c=y_{3}\frac{u}{\Lambda}$.\\
Case II: We obtain a zero in the $Y^{''}_{13}$ position of the Yukawa coupling matrix from Eq.\eqref{eq1} with a slight alteration in the assignment of vev allignment of the flavons as such:
$<\chi>=(u,u,u)$, $<\chi^{'}>=(u,u,u)$ and $<\chi^{''}>=(0,-u,u)$.
The Yukawa coupling matrix takes the form:
\begin{equation}
Y^{''} = \left[\begin{array}{ccc}
a & b & 0 \\
a & b & -c\\
a & b & c\\
\end{array}\right]\\.\label{Y13}
\end{equation}\\
Case III: Again with the choice of vev allignments given by:
$<\chi>=(u,u,u)$, $<\chi^{'}>=(0,-u,u)$ and $<\chi^{''}>=(u,u,u)$, the term in the $Y^{'''}_{22}$ position turns out to be zero. Thus, the Yukawa coupling matrix in this case is expressed as:
\begin{equation}
Y^{'''} = \left[\begin{array}{ccc}
a & b & c \\
a & 0 & c\\
a & -b & c\\
\end{array}\right]\\.\label{Y22}
\end{equation}\\
Considering these three cases for the Yukawa coupling matrices, we carry out our analysis in various sector. Incorporating the three cases of Yukawa coupling matrix in Eq.\ref{eq7}, we obtain a $\mu-\tau$ symmetry for Case I and Case II. A broken $\mu-\tau$ symmetry is obatined naturally only in Case III. Thus, we study the phenomenology for Case III, as it is the only viable stucture of one zero texture Yukawa coupling matrix in our model. The elements of the light neutrino mass matrix incorporating Case III are as follows:
\begin{equation}
	M^{\nu}_{11}= 1/32\pi^{2}\big[a^{2}M_{1}[L_{1}(m^{2}_{\eta_{R}^{0}}) - L_{1}(m^{2}_{\eta_{I}^{0}})]+b^{2}M_{2}[L_{2}(m^{2}_{\eta_{R}^{0}}) - L_{2}(m^{2}_{\eta_{I}^{0}})]+c^{2}M_{3}[L_{3}(m^{2}_{\eta_{R}^{0}}) - L_{3}(m^{2}_{\eta_{I}^{0}})]\big]
\end{equation}
\begin{equation}
M^{\nu}_{12}= 1/32\pi^{2}\big[a^{2}M_{1}[L_{1}(m^{2}_{\eta_{R}^{0}}) - L_{1}(m^{2}_{\eta_{I}^{0}})]+c^{2}M_{3}[L_{3}(m^{2}_{\eta_{R}^{0}}) - L_{3}(m^{2}_{\eta_{I}^{0}})]\big]
\end{equation}
\begin{equation}
M^{\nu}_{13}= 1/32\pi^{2}\big[a^{2}M_{1}[L_{1}(m^{2}_{\eta_{R}^{0}}) - L_{1}(m^{2}_{\eta_{I}^{0}})]-b^{2}M_{2}[L_{2}(m^{2}_{\eta_{R}^{0}}) - L_{2}(m^{2}_{\eta_{I}^{0}})]+c^{2}M_{3}[L_{3}(m^{2}_{\eta_{R}^{0}}) - L_{3}(m^{2}_{\eta_{I}^{0}})]\big]
\end{equation}
\begin{equation}
M^{\nu}_{22}= 1/32\pi^{2}\big[a^{2}M_{1}[L_{1}(m^{2}_{\eta_{R}^{0}}) - L_{1}(m^{2}_{\eta_{I}^{0}})]+c^{2}M_{3}[L_{3}(m^{2}_{\eta_{R}^{0}}) - L_{3}(m^{2}_{\eta_{I}^{0}})]\big]
\end{equation}
\begin{equation}
M^{\nu}_{23}= 1/32\pi^{2}\big[a^{2}M_{1}[L_{1}(m^{2}_{\eta_{R}^{0}}) - L_{1}(m^{2}_{\eta_{I}^{0}})]+c^{2}M_{3}[L_{3}(m^{2}_{\eta_{R}^{0}}) - L_{3}(m^{2}_{\eta_{I}^{0}})]\big]
\end{equation}
\begin{equation}
M^{\nu}_{33}= 1/32\pi^{2}\big[a^{2}M_{1}[L_{1}(m^{2}_{\eta_{R}^{0}}) - L_{1}(m^{2}_{\eta_{I}^{0}})]-b^{2}M_{2}[L_{2}(m^{2}_{\eta_{R}^{0}}) - L_{2}(m^{2}_{\eta_{I}^{0}})]+c^{2}M_{3}[L_{3}(m^{2}_{\eta_{R}^{0}}) - L_{3}(m^{2}_{\eta_{I}^{0}})]\big]
\end{equation}
	\section{Constraints on the model}\label{sec4}
	\subsection{Neutrinoless double beta decay}\label{A}
Analysing the neutrino phenomenology of the allowed Yukawa coupling matrix structure(Case III) in our work, we therefore, calculate the effective mass of the active neutrinos($m_{\beta\beta}$). The experimental technique of detecting the Majorana neutrino mass(which is a combination of the neutrino mixing matrix and the neutrino mass eigenstates) i.e. neutrinoless double beta decay($0\nu\beta\beta$)\cite{Mohapatra:1986su,Giunti:2004vv,Barry:2013xxa} includes some well known experiments related to it such as KamLAND-Zen\cite{Kamland2,kamland}, GERDA\cite{gerda,GERDA2}, KATRIN\cite{katrin2,KATRIN}. Its existence can be associated with the Majorana neutrinos. The expression for $m_{\beta\beta}$ is given by:
\begin{equation}
| m_{\beta\beta}|= \sum_{k=1}^3 m_{k} U_{ek}^{2}\label{eq:12}
\end{equation}
where, $\ U_{ek}^{2}$ are the elements of the neutrino mixing matrix with $k$ holding up the generation index. This eq.\eqref{eq:12} can be further expressed as,
\begin{equation}
| m_{\beta\beta}|= | m_{1} U_{ee}^{2} +  m_{2} U_{e\nu}^{2} + m_{3} U_{e\tau}^{2}|.
\end{equation} 
Calculation of the effective mass has a vital part in determining the possibility of the light neutrino parameters of a model to hint towards the sensitivity of the ongoing as well as future experiments.  
	\subsection{Lepton Flavor Violation(LFV)}\label{B}
	We estimate the viability of the model on basis of various lepton violating processes such as $l_{\alpha}\rightarrow l_{\beta}\gamma$,$l_{\alpha}\rightarrow 3 l_{\beta}$ and $\mu-e$ conversion in nuclei\cite{LFV,LFV2}. The most robust bound on the models mainly come from the muon decay experiments and the limits on this decay process i.e. $\rm BR(l_{\alpha} \rightarrow l_{\beta}\gamma)<4.2 \times 10^{-13}$ is set by the MEG collaboration\cite{TheMEG}. Future experiment MEG II may further improve this bound to a more precised one. SINDRUM experiment\cite{Perrevoort:2018cqi} gives a bound on $\rm BR(l_{\alpha} \rightarrow 3l_{\beta})$ which is  $\rm BR(l_{\alpha} \rightarrow 3l_{\beta}) <10^{-12}$. Interestingly a 4 orders improve on the magnitude of the current bound can be expected from the future Mu3e experiment. In case of the $\mu-e$ conversion of muonic atom, the experiments which essentially focuses on it are DeeMe\cite{DeeMee}, Mu2e\cite{Mu2e}, COMET\cite{comet} and PRIME\cite{prime}. The sensitivity on the limits produced from these experiments range from $10^{-14}$ to $10^{-18}$. A possibility of improving the current limits on $\tau$  in the near future is given by the LHC collaboration, as well as by B-factories such as Belle II\cite{Belle2}.

	\begin{figure}
		\includegraphics[width=0.3\textwidth]{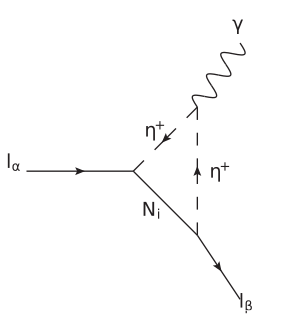}
		\includegraphics[width=0.3\textwidth]{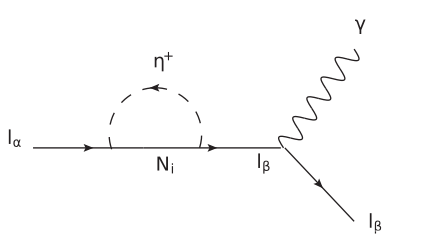}
		\includegraphics[width=0.3\textwidth]{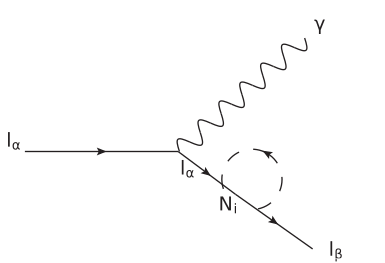}\\
		\caption{ The 1-loop Feynman diagrams depicting the decay of $l_{\alpha}\longrightarrow l_{\beta}\gamma$ \cite{Takashi}.} \label{fig4}
	\end{figure}
	We now discuss the analytical results of branching ratios of different LFV processes such as  $l_{\alpha}\rightarrow l_{\beta}\gamma$,$l_{\alpha}\rightarrow 3 l_{\beta}$ and $\mu-e$ conversion in nuclei in case of the scotogenic model.
	
	The branching ratio of $l_{\alpha}\rightarrow l_{\beta}\gamma$ for radiative lepton decay is given by\cite{Takashi}:
	
	\begin{equation}
	\rm BR(l_{\alpha}\rightarrow l_{\beta}\gamma)=\frac{3(4\pi^{3})\alpha_{em}}{4G_{F}^{2}}|A_{D}|^{2} BR(l_{\alpha} \rightarrow l_{\beta}\nu_{\alpha}\bar{\nu_{\beta}}).
	\end{equation}
Where, $\rm G_{F}$ is the Fermi constant and $\rm \alpha_{em}=\frac{e^{2}}{4\pi}$ is the electromagnetic fine structure constant and e denoting the electromagnetic coupling. The dipole form factor $\rm A_{D}$ is expressed as:
	
	\begin{equation}
	A_{D}=\sum_{i=1}^{3} \frac{Y_{i\beta}^{*} Y_{i\alpha}}{2(4\pi)^{2}} \frac{1}{m_{\eta^{+}}^{2}} F_{2}(\rho_{i})
	\end{equation}
	
	with $\rho_{i}$ being defined as $\rho_{i}=\frac{M_{i}^{2}}{m_{\eta^{+}}^{2}}$ and $F_{2}(x)$ is the loop function\cite{Takashi,LS}.
	
The branching ratio for three body decay process like $l_{\alpha}\rightarrow 3 l_{\beta}$\cite{Takashi} is as follows:
	\begin{equation}
	\begin{aligned}
	\rm BR(l_{\alpha}\rightarrow 3 l_{\beta})=\frac{3(4\pi^{2})\alpha_{em}^{2}}{8G_{F}^{2}} \bigg[|A_{ND}|^{2}+|A_{D}|^{2}\bigg(\frac{16}{3}\log\bigg(\frac{m_{\alpha}}{m_{\beta}}\bigg)-\frac{22}{3}\bigg)\\ 
	\rm +\frac{1}{6}|B|^{2}+\bigg(-2A_{ND} A_{D}^{*}+\frac{1}{3}A_{ND} B^{*}-\frac{2}{3}A_{D}B^{*}+h.c\bigg)\bigg] \\ \rm 
	\times BR(l_{\alpha} \rightarrow l_{\beta}\nu_{\alpha}\bar{\nu_{\beta}}).
	\end{aligned}
	\end{equation}

Considering $m_{\beta}<< m_{\alpha}$ only in the logarithmic term so that the appearance of an infrared divergence is refrained. The form factor $A_{D}$ is generated by dipole photon penguins and the other form factor $A_{ND}$ is  given by:
	
	\begin{equation}
	A_{ND}=\sum_{i=1}^{3} \frac{Y_{i\beta}^{*} Y_{i\alpha}}{6(4\pi)^{2}} \frac{1}{m_{\eta^{+}}^{2}} G_{2}(\rho_{i}).
	\end{equation}
	
	$\rm A_{ND}$ is generated by non-dipole photon penguins, whereas B, induced by box diagrams is given by-
	\begin{equation}
	\rm e^{2}B=\frac{1}{(4\pi)^{2}m_{\eta^{+}}^{2}}\sum_{i,j=1}^{3} \bigg[\frac{1}{2}D_{1}(\rho_{i},\rho_{j})Y_{j\beta}^{*}Y_{j\beta}Y_{i\beta}^{*}Y_{i\alpha}+\sqrt{\rho_{i}\rho_{j}}D_{2}(\rho_{i},\rho_{j})Y_{j\beta}^{*}Y_{j\beta}^{*}Y_{i\beta}Y_{i\alpha}\bigg].
	\end{equation}
	
	For the expressions of the functions $\rm G_{2}(x)$, $\rm D_{1}(x,y)$ and $\rm D_{2}(x,y)$ one can refer to \cite{Takashi,LS}. The e Z-boson penguin contributions are negligible as they
	are suppressed by charged lepton masses in this model. Also the contribution from Higgs-penguin are not considered as they too are suppressed.
	\begin{figure}
		\includegraphics[width=0.3\textwidth]{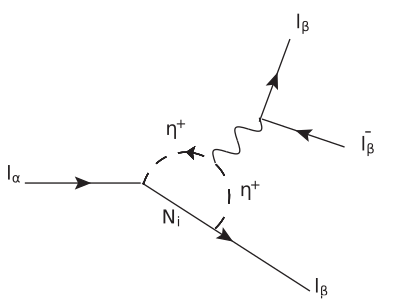}
		\includegraphics[width=0.3\textwidth]{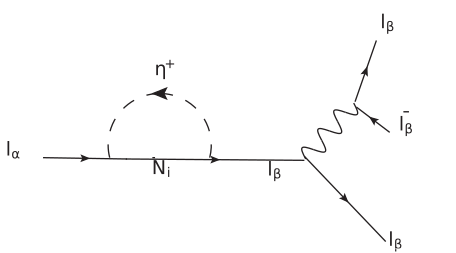}
		\includegraphics[width=0.3\textwidth]{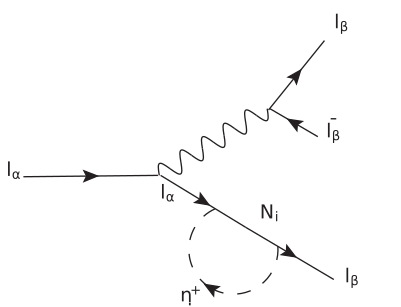}\\
		\caption{ The penguin contributions to $l_{\alpha}\longrightarrow l_{\beta}\gamma$, where the wavy lines depicts either a Z-boson or a photon \cite{Takashi}.} \label{fig5}
	\end{figure}
We can express the $\mu-e$ conversion rate, normalized
	to the muon capture rate by:
	\begin{equation}
	\begin{aligned}
	\rm CR(\mu-e,Nucleus)=\frac{p_{e}E_{e}m_{\mu}^{3}G_{F}^{2}\alpha_{em}^{3}Z_{eff}^{4}F_{p}^{2}}{8\pi^{2}Z \Gamma_{capt}}\times \bigg[|(Z+N)(g_{LV}^{(0)}+g_{LS}^{(0)})+(Z-N)(g_{LV}^{(1)}+g_{LS}^{(1)})|^{2}\\ \rm +|(Z+N)(g_{RV}^{(0)}+g_{RS}^{(0)})+(Z-N)(g_{RV}^{(1)}+g_{RS}^{(1)})|^{2}\bigg] .
	\end{aligned}
	\end{equation}
	
	Here, Z is the number of protons and N is the number of protons. $\rm Z_{eff}$ is the effective
	atomic charge, $F_{p}$ is the nuclear matrix element and $\rm \Gamma_{capt}$ is the total muon capture rate. Again, $\rm p_{e}$ and $E_{e}$ are the momentum and energy of electron respectively. We represent $g_{XK}^{(0)}$ and $g_{XK}^{(1)}$ (X=L,R and K=S,V)by the equations:
	\begin{equation}
	g_{XK}^{(0)}=\frac{1}{2} \sum_{q=u,d,s}\bigg(g_{XK}(q) G_{K}^{q,p}+g_{XK}(q) G_{K}^{q,n}\bigg)
	\end{equation}
	\begin{equation}
	g_{XK}^{(1)}=\frac{1}{2} \sum_{q=u,d,s}\bigg(g_{XK}(q) G_{K}^{q,p}-g_{XK}(q) G_{K}^{q,n}\bigg).
	\end{equation}
	
The contribution from the effective couplings $g_{XK}(q)$ in scotogenic model are given by:
	\begin{equation}
	g_{LV}(q) \approx g_{LV}^{\gamma}(q) 
	\end{equation} 
	\begin{equation}
	g_{RV}(q) = g_{LV}(q)|_{L\leftrightarrow R}
	\end{equation}
	\begin{equation}
	g_{LS}(q) \approx 0
	\end{equation}
	\begin{equation}
	g_{RS}(q) \approx 0
	\end{equation}
	
here, the photon penguins contribution is given by $\rm g_{LV}^{\gamma}(q)$. Now the effective coupling can be expressed as:
	\begin{equation}
	g_{LV}^{\gamma}(q) =\frac{\sqrt{2}}{G_{F}} e^{2} Q_{p} (A_{ND}-A_{D}).
	\end{equation}
	
	where, $Q_p$ is the electric charge of the corresponding quark.
	
	\subsection{Baryon asymmetry of the Universe(BAU)}\label{C}
	It is known that baryon asymmetry of the Universe can be produced via the mechanism of leptogenesis\cite{leptogenesis} for the out of equilibrium decay of $N_{1}\rightarrow l\eta, \bar{l}\eta^{*}$. In our work, we have analyzed the observed baryogenesis for a flavor symmetric scotogenic model in order to check its feasibility in terms of cosmological aspect. In the scotogenic model framework, a crucial result has been discussed in many literatures\cite{Hugle:2018qbw} which highlights the existence of a lower bound for the lightest of the RHNs($M_{1}$) i.e about 10TeV considering the vanilla leptogenesis scenario\cite{Hugle:2018qbw,Borah:2018rca}. The lepton asymmetry generated is only due to the decay of $N_{1}$ as the decay of $N_{2}$ and $N_{3}$ are supressed because of strong washout effects produced by $N_{1}$ or $N_{2}$ and $N_{3}$ mediated interactions\cite{Borah:2018rca}. This occurs as a we consider a mass heirarchy of the RHN masses, i.e $M_{1} << M_{2}, M_{3}$. We consider the RHN masses as $M_{1}= 10^{4}-10^{5}$ GeV, $M_{2}= 10^{6}-10^{7}$ GeV and $M_{3}= 5\times 10^{7}-10^{8}$ GeV. Also considering $m_{\eta^{0}_{R}}= 450-750$ GeV and the lightest neutrino mass $m_{l}= 10^{-12}-10^{-10}$ eV, corresponds to the weak washout regime. The decay parameter governs the distinction between weak and strong washout regime which is further an essential component in calculating leptogenesis. Its relation is given by: 
	\begin{equation}
	K_{N_{1}}= \frac{\Gamma_{1}}{H(z=1)},
	\end{equation}
	where, $\Gamma_{1}$ is the total $N_{1}$ decay width, $H$ is the Hubble parameter, $z= \frac{M_{1}}{T}$ and $T$ is the temperature of the photon bath. Again, we can express H as:
	\begin{equation}\label{eq:1}
	H = \sqrt\frac{8\pi^{3}g_{*}}{90}\dfrac{T^{2}}{M_{Pl}}.
	\end{equation} 
	In Eq.\eqref{eq:1},the effective number of relativistic degrees of freedom is given by  $g_{*}$ and $M_{Pl}\simeq 1.22\times 10^{19}$ GeV stands for the Planck mass. The constrained Yukawa couplings calculated from the model have a vital role in the decay rate equation for $N_{1}$ which is given by,
	\begin{equation}
	\Gamma_{1}= \frac{M_{1}}{8\pi}(Y^{\dagger}Y)_{11}\left[1- \Big(\frac{m_{\eta^{0}_{R}}}{M_{1}}\Big)^{2}\right]^{2}= \frac{M_{1}}{8\pi}(Y^{\dagger}Y)_{11}(1-\eta_{1})^{2}
	\end{equation}

The CP asymmetry parameter $\epsilon_{1}$ in its simplified form is given by, 
	\begin{equation}
	\epsilon_{1}= \frac{1}{8\pi(Y^{\dagger}Y)_{11}}\sum_{j\ne 1}Im[(Y^{\dagger}Y)^{2}]_{1j}\left[ f(r_{j1},\eta_{1})- \frac{\sqrt{r_{j1}}}{r_{j1}-1}(1-\eta_{1})^{2}\right],
	\end{equation}
	where,
	\begin{equation}
	f(r_{j1},\eta_{1})= \sqrt{r_{j1}}\left[1+ \frac{(1-2\eta_{1}+r_{j1})}{(1-\eta_{1})^{2}}  ln(\frac{r_{j1}-\eta_{1}^{2}}{1-2\eta_{1}+r_{j1}})\right],
	\end{equation}
	and
	$r_{j1}= \big(\frac{M_{j}}{M_{1}}\big)^{2}$, $\eta_{1}\equiv \big(\frac{m_{\eta^{0}_{R}}}{M_{1}}\big)^{2}$.\\
	The Boltzmann equations for the number densities of $N_{1}$ and $N_{B-L}$ are as follows \cite{Davidson:2002qv},
	\begin{equation}\label{eq:3}
	\frac{dn_{N_{1}}}{dz}= -D_{1}(n_{N_{1}} - n_{N_{1}}^{eq}),
	\end{equation}
	\begin{equation}\label{eq:4}
	\frac{dn_{B-L}}{dz}= -\epsilon_{1}D_{1}(n_{N_{1}} - n_{N_{1}}^{eq})- W_{1}n_{B-L},
	\end{equation}
	respectively. Here, $n_{N_{1}}^{eq}= \frac{z^{2}}{2}K_{2}(z)$ is the equilibrium number density of $N_{1}$, where $K_{i}(z)$ is the modified  Bessel function of $i^{th}$ type and
	\begin{equation}
	D_{1}\equiv \frac{\Gamma_{1}}{Hz} = K_{N_{1}}z\frac{K_{1}(z)}{K_{2}(z)}
	\end{equation} 
	gives the measure of the total decay rate with respect to the Hubble rate, and $ W_{1}= \frac{\Gamma_{W}}{Hz}$ is the total washout rate. We have $W_{1}= W_{1D}+W_{\Delta L=2}$, viz the summation of the washout due to inverse decays $l\eta,\bar{l}\eta^{*}\rightarrow N_{1}$ ($W_{1D}= \frac{1}{4}K_{N_{1}}z^{3}K_{1}(z)$) and the washout due to the $\Delta L= 2$ scatterings $l\eta \leftrightarrow \bar{l}\eta^{*},ll \leftrightarrow \eta^{*}\eta^{*}$ which is given by,
	\begin{equation}
	W_{\Delta L=2} \simeq \dfrac{18\sqrt{10}M_{Pl}}{\pi^{4}g_{l}\sqrt{g_{*}}z^{2}v^{4}}(\frac{2\pi^{2}}{\lambda_{5}})^{2}M_{1}\bar{m_{\varsigma}}^{2}.
	\end{equation}
	Here, $g_{l}$ stands for the internal degrees of freedom for the SM leptons, and $\bar{m_{\varsigma}}$ i.e. the effective neutrino mass parameter is defined as:
	\begin{equation}
	\bar{m_{\varsigma}}^{2} \simeq 4\varsigma_{1}^{2}m_{1}^{2} + \varsigma_{2}m_{2}^{2} +\varsigma_{3}^{2}m_{3}^{2},
	\end{equation}
	where, $m_{i}^{,}s$ are the light neutrino mass eigenvalues and $ \varsigma_{k}$ is expressed as:
	\begin{equation}
	\varsigma_{k} = \Big(\frac{M^{2}_{\textit{k}}}{8(m_{\eta_{R}^{0}}^{2}-m_{\eta_{I}^{0}}^{2})}[L_{k}(m^{2}_{\eta_{R}^{0}}) - L_{k}(m^{2}_{\eta_{I}^{0}})]\Big)^{-1}	
	\end{equation} 
	Now by numerical analysis of Eq.\eqref{eq:3} and Eq.\eqref{eq:4} before the sphaleron freeze-out, we find the final B-L asymmetry $n_{B-L}^{f}$. This is further converted into the baryon-to-photon ratio given by:
	\begin{equation}\label{eq:5}
	n_{B}= \frac{3}{4}\frac{g_{*}^{0}}{g_{*}}a_{sph}n_{B-L}^{f}\simeq 9.2\times 10^{-3}n_{B-L}^{f},
	\end{equation} 
	At the time of final lepton asymmetry production, the effective relativistic degrees of freedom is given by $g_{*}= 110.75$, whereas at the recombination epoch the effective degrees of freedom is  $g_{*}^{0}= \frac{43}{11}$. The sphaleron conversion factor is given by $a_{sph}=\frac{8}{23}$. The cosmological constraint on the observed BAU($n_{B}^{obs}$) is found to be $(6.04\pm0.08)\times 10^{-10}$\cite{Aghanim:2018eyx} as given by Planck limit 2018. In our analysis, the free parameters chosen are successful in satisfying the Planck bound for BAU. As we have also studied LFV for this framework, we look forward to satisfying the bounds on it while simultaneously generating the observed BAU. The quartic coupling, $\lambda_{5}$ is a significant parameter which can be fine tuned so that the constraints on the model are obeyed. In our work, we have taken it in the range $10^{-3}- 1$ and carried out our numerical analysis.

	\subsection{Scalar dark matter}\label{D}
As the expansion rate of the Universe becomes more than the pair annihilation rate, the particles decouples from the cosmic plasma, thereby losing its equilibrium state.
	On solving the Boltzmann equation \cite{Scherrer:1985zt,Kolb:1990vq}, we can obtain the relic densities of the thermally produced dark matter candidates:
	\begin{equation}
	\dot{n}_{DM} + 3Hn_{DM} = -<\sigma v> (n^{2}_{DM}- (n^{eq}_{DM})^{2}),
	\end{equation}
	where, the number density of the dark matter candidate and the number density of the dark matter candidate in thermal equilibrium is denoted by $n_{DM}$ and $n_{DM}^{eq}$ respectively .% The numerical solution of the Boltzmann equation in terms of partial wave expansion, $<\sigma v> = a + bv^{2}$ is of the form,
%	\begin{equation}
%	\Omega h^{2}\approx \frac{1.04 \times 10^{9} x_{f}}{M_{Pl}\sqrt{g_{*}}(a + 3b/x_{f})},
%	\end{equation}
%	where, $x_{f}= \frac{m_{DM}}{T_{f}}$, $T_{f}$ is the freeze-out temperature, also  $v^{2}\simeq\frac{6}{x_{f}}$, $m_{DM}$ is the mass of dark matter, $g_{*}$ is the number of relativistic degrees of freedom at the time of freeze-out, and $M_{Pl} \approx 1.22 \times 10^{19}$ GeV is the Planck mass. 
A simplified analytical form for the approximation of DM relic abundance is expressed as \cite{Jungman:1995df},
	\begin{equation}
	\Omega h^{2} \approx \dfrac{3 \times 10^{-27} cm^{3} s^{-1}}{<\sigma v>}
	\end{equation}
	The corresponding thermal averaged annihilation cross section is further given by\cite{Gondolo:1990dk};
	\begin{equation}
	<\sigma v> = \dfrac{1}{8m_{DM}^{4}TK_{2}^{2}(m_{DM}/T)}\int_{4m_{DM}^{2}}^{\infty}\sigma (s-4m_{DM}^{2})\sqrt{s}K_{1}(\sqrt{s}/T) ds ,
	\end{equation}
	where, $K_{1}$ and $K_{2}$ are the modified Bessel functions, $m_{DM}$ is the mass of dark matter candidate and $T$ is the temperature.
	In this $A_{4}\otimes Z_{4}$ realisation of the scotogenic model, the lightest of the neutral component of the scalar doublet $\eta$ , i.e,  $\eta^{0}$ is considered to be the dark matter candidate. \cite{LopezHonorez:2006gr,Ahriche:2017iar,Deshpande:1977rw,Cirelli:2005uq,Barbieri:2006dq,Ma:2006wm,Hambye:2009pw,Dolle:2009fn,Honorez:2010re,Gustafsson:2012aj,Borah:2017dfn,Goudelis:2013uca,Arhrib:2013ela,Bhattacharya:2019fgs,Borah:2019aeq}. 
	% The low mass regime, wherein the DM mass $m_{\eta^{0}}\le M_{W}$, DM annihilation into the SM fermions through $s$-channel Higgs mediation is dominant. As mentioned in the literature \cite{IHDM8}, the possible DM annihilations $\eta^{0}\eta^{0} \rightarrow WW^{*}\rightarrow Wf\bar{f^{/}}$ can also be significant in the low mass regime. Also, the co-annihilations of $\eta^{0}_{R}, \eta^{\pm}~ \text {and}~ \eta^{0}_{I}$ subjective to the mass differences $m_{\eta^{\pm}}-m_{\eta^{0}_{R}}\equiv \Delta M_{\eta^{\pm}}, m_{\eta^{0}_{I}}- m_{\eta^{0}} \equiv \Delta M_{\eta^{0}_{I}} $ can also play a role in the generation of relic abundance of DM. A comprehensive study on this type of co-annihilations are being discussed in \cite{relic2,relic3,relic4}.
	The effective cross-section is given by\cite{Griest:1990kh}: 
	\begin{equation}
	\sigma_{eff} = \sum_{i,j}^{N} <\sigma_{ij} v> \frac{g_{i} g_{j}}{g_{eff}^{2}}(1 + \Delta_{i})^{3/2}(1 + \Delta_{j})^{3/2} e^{(-x_{f}(\Delta_{i} + \Delta_{j}))},
	\end{equation}
	with, $\Delta_{i}= \frac{m_{i}- m_{DM}}{m_{DM}}$ and $
	g_{eff}= \sum_{i=1}^{N} g_{i}(1+\Delta_{i})^{3/2} e^{-x_{f}\Delta_{i}}.$
	
	Here, $ m_{i}$ is the mass of the heavier inert Higgs doublet. Thus, we can express the thermally averaged cross section by:
	\begin{equation}
	<\sigma_{ij} v> = \dfrac{x_{f}}{8 m_{i}^{2} m_{j}^{2} m_{DM} K_{2}(\frac{m_{i}x_{f}}{m_{DM}}) K_{2} (\frac{m_{j} x_{f}}{m_{DM}})}\times \int_{(m_{i}+ m_{j})^{2}}^{\infty}  \sigma_{ij}(s- 2(m_{i}^{2}+ m_{j}^{2})) \sqrt{s} K_{1} \big(\frac{\sqrt{s}x_{f}}{m_{DM}}\big)ds.
	\end{equation}
	
\begin{figure}[H]
		\begin{center}
			\includegraphics[width=0.4\textwidth]{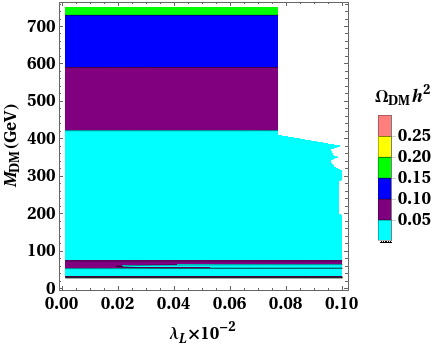}
			\includegraphics[width=0.4\textwidth]{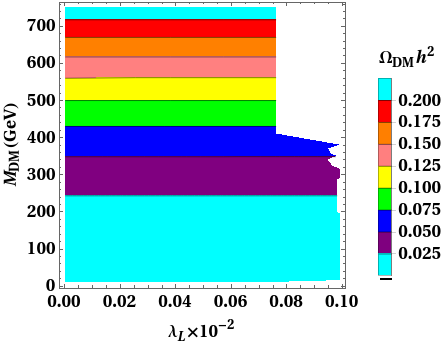}
		\end{center}
		\caption{ Contour plot between the DM-Higgs coupling $\lambda_{L}$ and DM mass $M_{DM}$ w.r.t the allowed space of relic abundance of DM. The left panel is for $\Delta M_{\eta^{\pm}}$ = $\Delta M_{\eta^{0}_{I}} = 8$ GeV and the right panel is for $\Delta M_{\eta^{\pm}}$ = $\Delta M_{\eta^{0}_{I}} = 0.9$ GeV }\label{DM}
	\end{figure}

\begin{figure}[H]
		\begin{center}
			\includegraphics[width=0.4\textwidth]{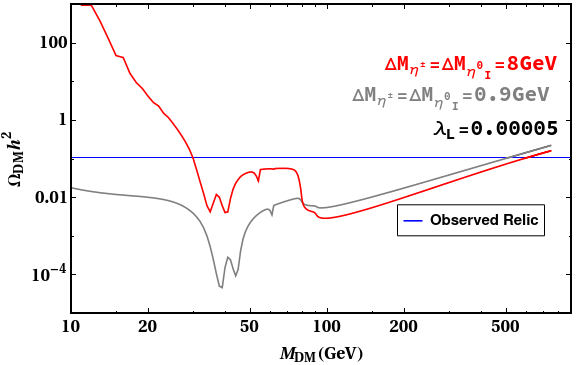}
		
		\end{center}
		\caption{ Variational plot between relic abundance of DM ($\Omega_{DM}h^{2}$) and DM mass($M_{DM}$) with benchmark value $\lambda_{L}=0.00005$ for two different values of mass splitting between inert scalars, i.e. $\Delta M_{\eta^{\pm}}$ = $\Delta M_{\eta^{0}_{I}} = 8$ GeV and $\Delta M_{\eta^{\pm}}$ = $\Delta M_{\eta^{0}_{I}} = 0.9$ GeV. }\label{DM1}
	\end{figure}

	 In our work, we have shown the relic abundance for a certain range of dark matter, i.e $ M_{DM}= 450-750$ GeV by the usage of {\tt MicrOmega 5.0.4}\cite{Belanger:2018ccd}. 
	Due to the choice of RHN masses being heavier than the DM mass, its influence in the dark matter sector is negligible, i.e. it doesn't alter the relic abundance generated for the lightest inert scalar.\\
	The parameters playing a crucial role in the generation of relic abundance is the DM-Higgs coupling ($\lambda_{L}$) and the inert scalar mass splittings. By appropriate choosing these parameters we can successfully obtain the correct relic abundance for DM mass, i.e. around 450-500 GeV. We have done a comparative study so as to show how crucial the mass splitting between the inert scalar can be for the production of observed relic abundance. We have chosen two cases, $\Delta M_{\eta^{\pm}}$ = $\Delta M_{\eta^{0}_{I}} = 8$ GeV and $\Delta M_{\eta^{\pm}}$ = $\Delta M_{\eta^{0}_{I}} = 0.9$ GeV respectively in Fig.\ref{DM1}, where we can observe that for small mass splitting the relic is achieved for $M_{DM}\sim 480$ GeV, whereas for large mass splitting we get the relic for DM mass above 500 GeV. From Fig.\ref{DM}, we have analysed the parameter space of $\lambda_{L}$ for two different values of scalar mass splittings which satisfy the Planck limit for relic abundance of DM. The left panel of Fig.\ref{DM} corresponds to $\Delta M_{\eta^{\pm}}$ = $\Delta M_{\eta^{0}_{I}} = 8$ GeV, where the DM mass satisying the observed relic abundance limit is above 580 GeV for the values of $\lambda_{L}$ upto $0.08\times 10^{-2}$. Whereas for $\Delta M_{\eta^{\pm}}$ = $\Delta M_{\eta^{0}_{I}} = 0.9$ GeV, relic abundance is satisfied for $M_{DM}\sim 490-500$ GeV for the same range of $\lambda_{L}$ as can be seen in the right panel of Fig.\ref{DM}.
	
	\section{Results and analysis}\label{sec7}
	We have discussed in Sec.\ref{sec3} an $A_{4}\otimes Z_{4}$ extension of the minimal Scotogenic model. By the choice of three different sets of vev alignment, we are able to show three distinct structures of Yukawa coupling matrices bearing a zero component in  one of the matrix elements. However, two of the structures can be discarded from the $\mu-\tau$ symmetry point of view. The first two cases of one zero texture are seen to take the form of $\mu-\tau$ symmetry when incorporated in the mass matrix of the model. We are therefore left with only one structure of one zero texture Yukawa coupling matrix which breaks the $\mu-\tau$ symmetry and thus is allowed in the model. 
	
	Also, a notable kind of parametrization known as the {\it Casas-Ibarra parametrization }\cite{Casas:2001sr} is used in our work to numerical obtain the numerical values of the model parameters. This also helps us in relating the Yukawa coupling with the light neutrino parameters. The 3-$\sigma$ values neutrino oscillation parameters are taken from the literature\cite{3sigma}.
	\begin{equation}
	Y= U\sqrt{M^{diag}_{\nu}}R^{\dagger}\sqrt{\Lambda},\label{eq:CI}
	\end{equation}
	where $R$ is a complex orthogonal matrix which obeys the condition $R^{T}R=1$. We, thus choose the orthogonal complex matrix $R$ as:
	\begin{equation}\label{eq:11}
	R=
	\begin{pmatrix}
	0 & 0 & 1\\
\cos Z & -\sin Z & 0 \\
	\sin Z & \cos Z & 0\\
	\end{pmatrix},
	\end{equation}
	which can be further expressed as:
		\begin{equation}\label{eq:111}
	R=
	\begin{pmatrix}
	0 & 0 & 1\\
	\theta & - \sqrt{1-\theta^{2}} & 0 \\
	\sqrt{1-\theta^{2}} & \theta & 0\\
	\end{pmatrix},
	\end{equation}
	where, $\theta= \cos Z$. The numerical value of $\theta$ is solved for the three different structures of the Yukawa coupling matrix.

Since Case I and Case II are discarded, we carry out our study only considering Case III.
\subsection{Case III:}
\begin{figure}
\begin{center}	
	\includegraphics[width=0.4\textwidth]{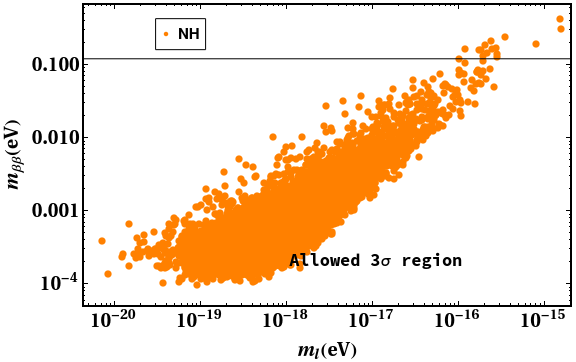}
	\includegraphics[width=0.4\textwidth]{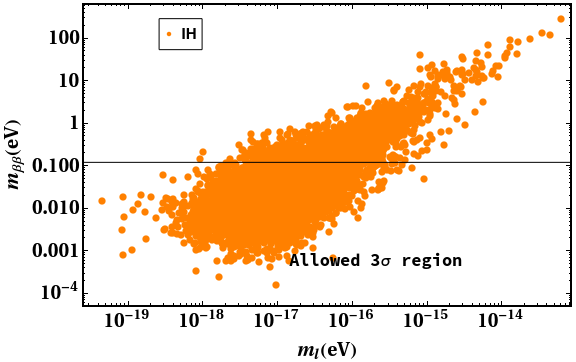}
	\caption{Variation of the lightest active neutrino mass($m_{l}$) with effective mass($m_{\beta\beta}$) in Case III for NH/IH. The KamLAND-Zen limit $m_{\beta\beta}(eV)\sim0.1(eV)$ is shown by the horizontal(black) line.}\label{eff22}
	\end{center}
\end{figure}

\begin{figure}[H]
	\begin{center}
	\includegraphics[width=0.3\textwidth]{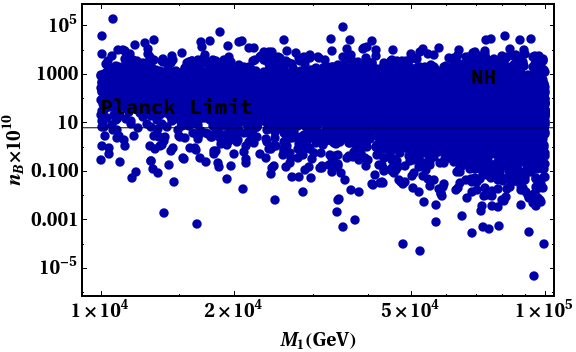}
	\includegraphics[width=0.3\textwidth]{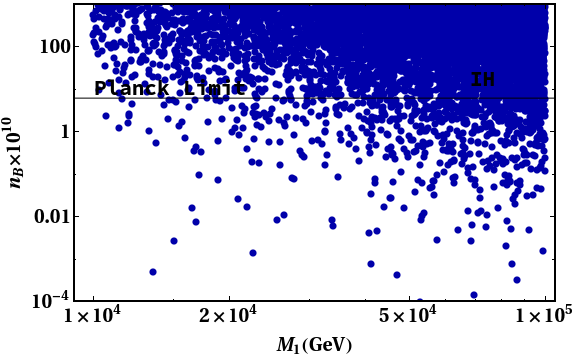}\\
	\includegraphics[width=0.3\textwidth]{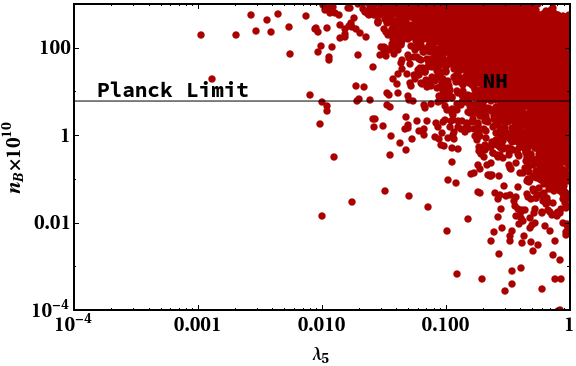}
	\includegraphics[width=0.3\textwidth]{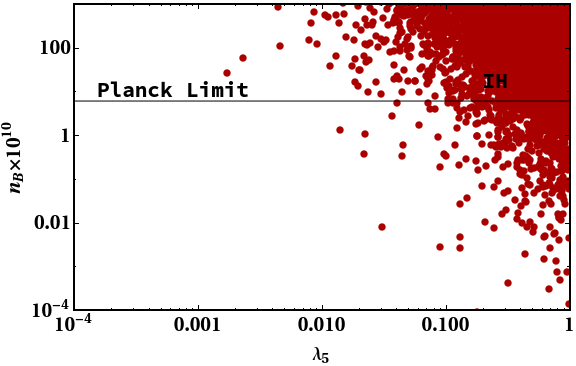}\\
	\includegraphics[width=0.3\textwidth]{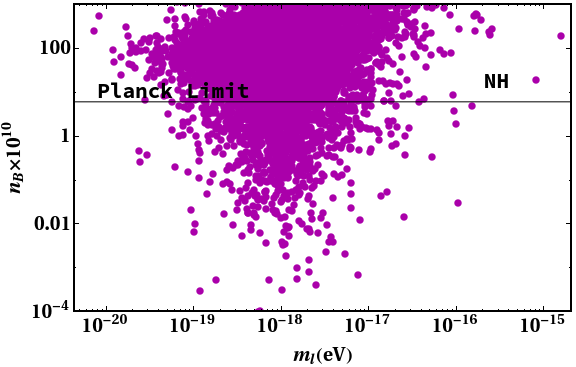}
	\includegraphics[width=0.3\textwidth]{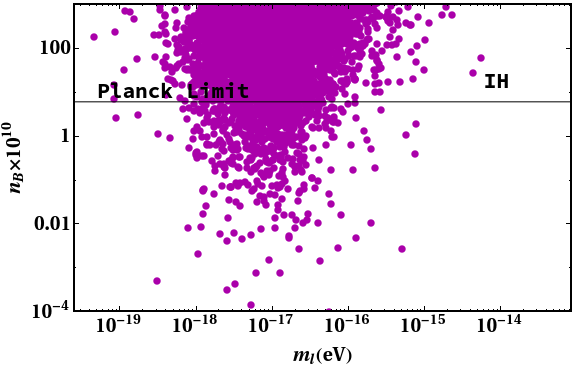}\\
	\includegraphics[width=0.3\textwidth]{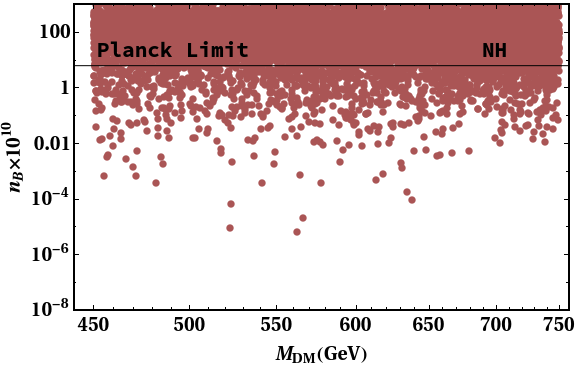}
	\includegraphics[width=0.3\textwidth]{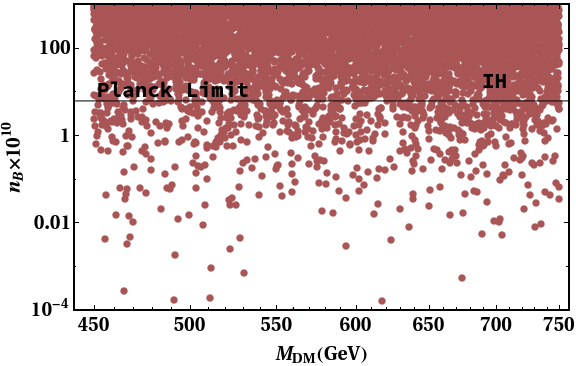}\\
	
		\caption{We showcase plots of baryon asymmetry w.r.t four parameters namely lightest RHN mass ($M_{1}$), quartic coupling ($\lambda_5$), lightest neutrino mass eigenvalue($m_{l}$) and dark matter mass($M_{DM}$)respectively for Case III. The black horizontal line gives the current Planck limit for BAU. }\label{bau22}
	\end{center}
\end{figure}

\begin{figure}[H]
	\begin{center}
		\includegraphics[width=0.4\textwidth]{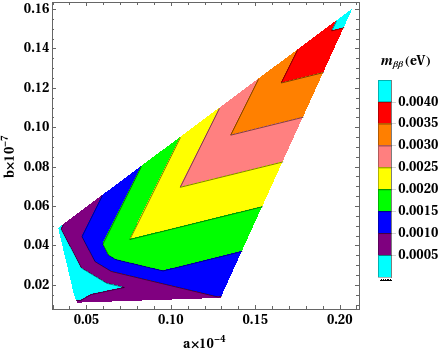}
		\includegraphics[width=0.4\textwidth]{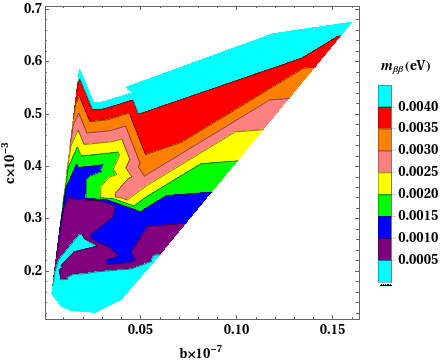}
		\caption{ Contour plot showing the parameter space of model parameters a,b and c w.r.t effective mass($m_{\beta\beta}$) for NH for Case III. }\label{C22}
		
	\end{center}
\end{figure}

\begin{figure}[H]
		\includegraphics[width=0.3\textwidth]{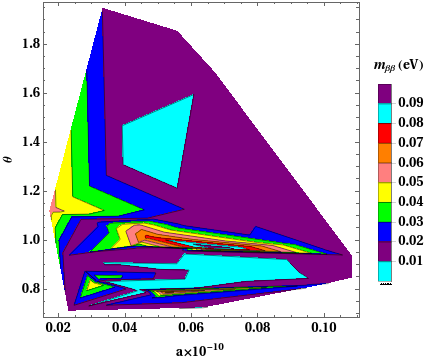}
		\includegraphics[width=0.3\textwidth]{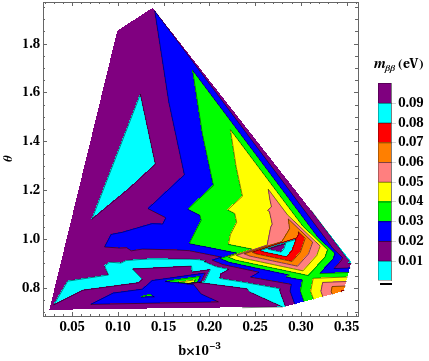}
		\includegraphics[width=0.3\textwidth]{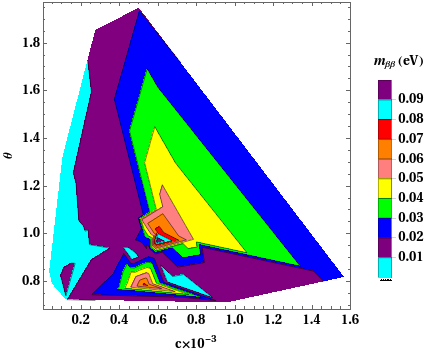}
		\caption{ Contour plot showing the parameter space of model parameters a,b,c and rotational angle $\theta$ w.r.t effective mass($m_{\beta\beta}$) for IH for Case III. }\label{C221}
		
\end{figure}

\begin{figure}[H]
	\begin{center}
		\includegraphics[width=0.3\textwidth]{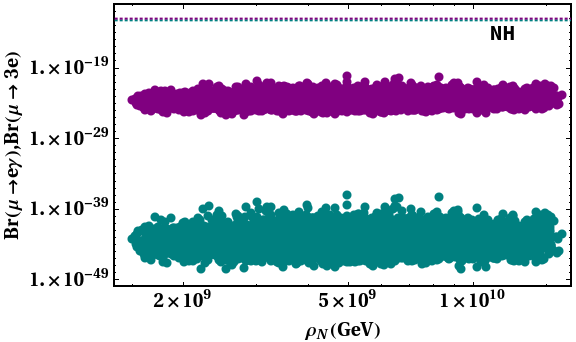}
		\includegraphics[width=0.3\textwidth]{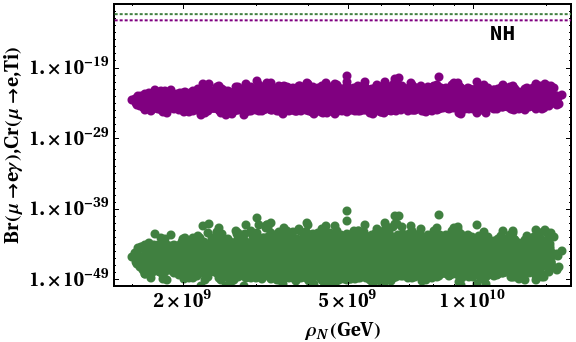}
		\includegraphics[width=0.3\textwidth]{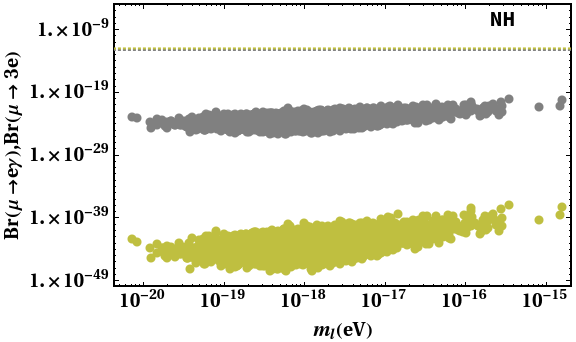}
		\caption{Variation of $Br(\mu\rightarrow e\gamma)$ and $Br(\mu\rightarrow 3e)$ with $\rho_{N}$( $\rho_{N}=(\frac{M_{N}}{m_{\eta^{+}}})^{2}$)(left panel), plot depicting $Br(\mu\rightarrow e\gamma)$ and $Cr(\mu\rightarrow e,Ti)$ as a function of $\rho_{N}$(middle) and $Br(\mu\rightarrow e\gamma)$ and $Br(\mu\rightarrow 3e)$ w.r.t the lightest neutrino mass eigenvalue($m_{l}$) is shown in the right panel in Case III for NH. The upper bounds are shown by the horizontal lines.}\label{lfv22}
		
	\end{center}
\end{figure}

\begin{figure}[H]
	\begin{center}
		\includegraphics[width=0.3\textwidth]{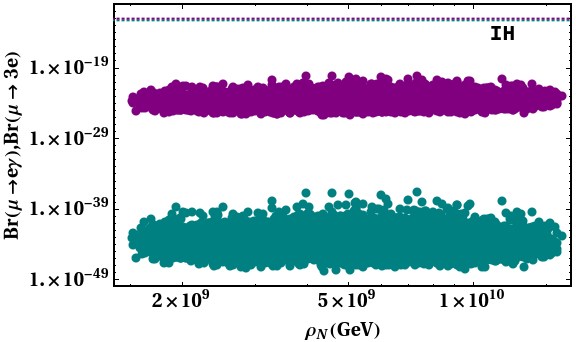}
		\includegraphics[width=0.3\textwidth]{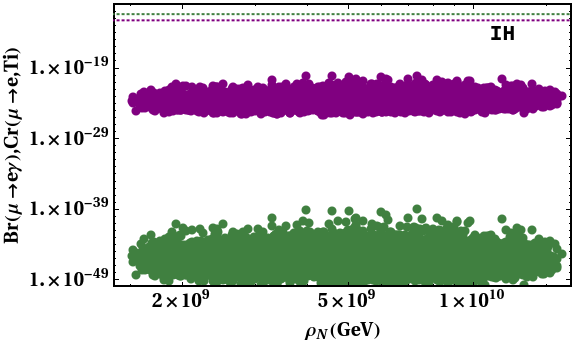}
		\includegraphics[width=0.3\textwidth]{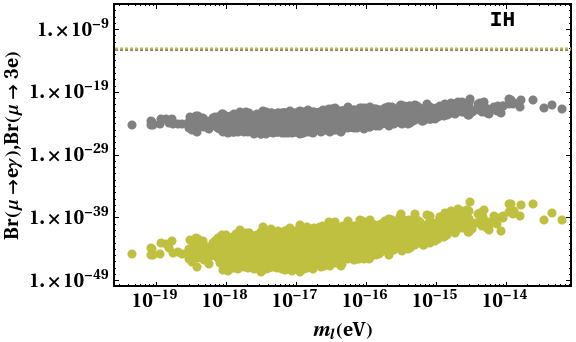}
		\caption{Variation of $Br(\mu\rightarrow e\gamma)$ and $Br(\mu\rightarrow 3e)$ with $\rho_{N}$( $\rho_{N}=(\frac{M_{N}}{m_{\eta^{+}}})^{2}$)(left panel), plot depicting $Br(\mu\rightarrow e\gamma)$ and $Cr(\mu\rightarrow e,Ti)$ as a function of $\rho_{N}$(middle) and $Br(\mu\rightarrow e\gamma)$ and $Br(\mu\rightarrow 3e)$ w.r.t the lightest neutrino mass eigenvalue($m_{l}$) is shown in the right panel in Case III for IH. The upper bounds are shown by the horizontal lines.}\label{lfv221}
		
	\end{center}
\end{figure}

\begin{figure}[H]
	\begin{center}
		\includegraphics[width=0.4\textwidth]{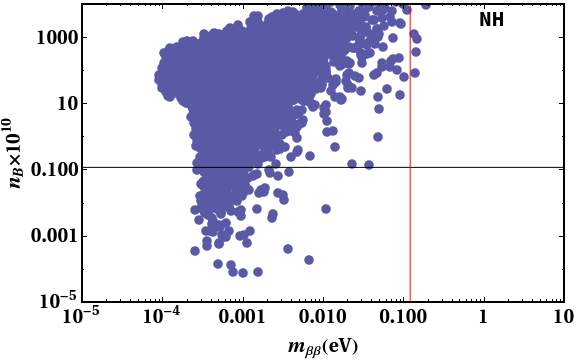}
		\includegraphics[width=0.4\textwidth]{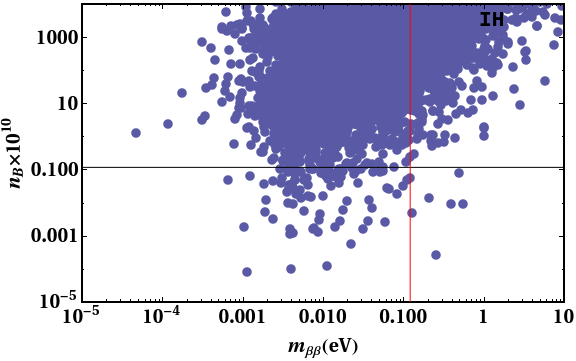}
		\caption{A co-relation plot of effective mass of active neutrinos($m_{\beta\beta}$) w.r.t the baryon asymmetry of the Universe($n_{B}$) is shown for Case III. The Planck limit for BAU is given by the horizontal line(black) and the KamLAND-Zen upper limit ($m_{\beta\beta}(eV)\sim0.1(eV)$) is depicted by the vertical line(red).}\label{nBmB22}
			\end{center}
\end{figure}

Again on interchanging the choice of vev alignment of the flavons $\chi$, $\chi^{'}$ and $\chi^{''}$, we can achieve a Yukawa coupling matrix with one zero element at the $Y^{'''}_{22}$ position as given in Eq.\ref{Y22}. As analysed earlier in the previous subsections, we follow a similar study for this particular structure of $Y^{'''}$. From Fig.\ref{eff22}, we see that in the plot for NH, all the points fall on the allowed region as per the KamLAND-Zen limit. However, in the right panel of Fig.\ref{eff22}, i.e. for IH, $m_{l}= 10^{-19}-10^{-16}$ eV is successful in generating the effective mass of active neutrinos in the allowed region. We show variation of different parameters vs baryon asymmetry of the Universe in Fig.\ref{bau22}. For NH, the mass range considered for $M_{1}$ satisfies the BAU limit given by Planck, however we have maximum points only in the region $M_{1}= 4\times10^{4}-10^{5}$ GeV in case of IH which generates the desired BAU. The parameter space of $\lambda_{5}$ satisfying the BAU constraint is between $10^{-1}-1$ for both NH and IH. In the third row of Fig.\ref{bau22}, we can conclude a definite range of lightest active neutrino which produces the desired BAU. For NH, $m_{l}= 10^{-19}-10^{-17}$ eV and for IH, $m_{l}= 10^{-18}-10^{-16}$ eV obeys the Planck limit for BAU. Again, considering the variation of $M_{DM}$ as a function of BAU, the entire range of DM mass, i.e. $M_{DM}=450-750$ GeV is seen to satisfy the Planck limit for BAU in case of both NH as well as IH. Unlike the other cases discussed above, here the orthogonal matrix R for NH has no variation in the matrix elements, i.e. the absolute value of all elements is found to be 1. Therefore, in Fig.\ref{C22}, contour plots of only a,b and c w.r.t $m_{\beta\beta}$ is shown. However, we obtain a different matrix for IH, thus we have shown plots considering its variation with other model parameters w.r.t $m_{\beta\beta}$ as can be seen in Fig.\ref{C221}. The allowed parameter space for the model parameters considered from the Fig\ref{C22} and Fig.\ref{C221} is shown in a tabular form in Tab.\ref{TAB5}. In case III, we can draw analysis from Fig.\ref{lfv22} and Fig.\ref{lfv221} that the variations of $\rho_{N}$ and $m_{l}$ w.r.t the branching ratios and conversion ratio is same for both NH and IH. Thus, we have $Br(\mu\rightarrow 3e)$ lies in the range $10^{-49}-10^{-39}$ and $Br(\mu\rightarrow e\gamma)\sim 10^{-26}-10^{-22}$ w.r.t $\rho_{N}$ and $m_{l}$. And the conversion ratio ranges from $Cr(\mu\rightarrow e,Ti)\sim10^{-49}-10^{-39}$. In our attempt to corelate the BAU and the effective mass of neutrinos $m_{\beta\beta}$, we have shown a corelation plot in Fig.\ref{nBmB22}. Here, we observe that for $m_{\beta\beta}= 10^{-4}-10^{-3}$eV the desired BAU is obtained in NH, whereas for IH, $m_{\beta\beta}= 10^{-2}-10^{-1}$eV is the allowed range generating correct BAU.

\begin{table}
	\begin{center}
		\begin{tabular}{|c|c|c|}
			
			\hline 
			Parameter& NH& IH \\ 
			\hline
			a &$0.05\times 10^{-8}-0.2\times 10^{-8}$& $0.02\times 10^{-10}-0.10\times 10^{-10}$ \\
			\hline 
			b &$0.01\times 10^{-7}-0.15\times 10^{-7}$ & $0.03\times 10^{-3}-0.35\times 10^{-3}$ \\
			\hline 
			c & $0.1\times 10^{-3}- 0.7\times 10^{-3}$& $0.1\times 10^{-3}- 1.6\times 10^{-3}$ \\
			\hline 
			$\theta$ &$1$&$0.7- 1.9$  \\
			\hline 	
		\end{tabular} 
	\end{center}
	\caption{Model parameters of the model and their respective parameter space satisfying effective mass of light neutrinos($m_{\beta\beta}$) for Case III.} \label{TAB5}
\end{table}


%\begin{figure}[H]
%	\begin{center}
%		\includegraphics[width=0.3\textwidth]{abNB22.png}
%		\includegraphics[width=0.3\textwidth]{bcNB22.png}
%		\caption{ Contour plot showing the parameter space of model parameters a,b and c w.r.t baryon asymmetry of the Universe($n_{B}$) for NH for Case III. }\label{C222}
		
%	\end{center}
%\end{figure}

%\begin{figure}[H]
%	\begin{center}
%		\includegraphics[width=0.3\textwidth]{aNBIH22.png}
%		\includegraphics[width=0.3\textwidth]{bnBIH22.png}
%		\includegraphics[width=0.3\textwidth]{cIH22.png}
%		\caption{ Contour plot showing the parameter space of model parameters a,b,c and rotational angle $\theta$ w.r.t baryon asymmetry of the Universe($n_{B}$) for IH for Case III. }\label{C223}
		
%	\end{center}
%\end{figure}


%\begin{table}[h]
%	\begin{center}[H]
%		\begin{tabular}{|c|c|c|c|}
%			
%			\hline 
%			Textures & $0\nu\beta\beta$ & Baryogenesis & LFV \\ 
%			\hline
%			Case I(NH/IH) &$\checkmark$($\checkmark $)& $\times$($\times$)&$\checkmark$($\checkmark$) \\
%			\hline 
%			Case II(NH/IH) &$\checkmark$($\checkmark$) &$\checkmark$($\checkmark$)&$\checkmark$($\checkmark$) \\
%			\hline 
%			Case III(NH/IH) & $\checkmark$($\checkmark$)& $\checkmark$($\checkmark$)&$\checkmark$($\checkmark$) \\
%			\hline 
		
%		\end{tabular} 
%	\end{center}
%	\caption{Allowed and disallowed phenomenologies for three cases of textures} \label{Texture}
%\end{table}
Analogous to the framework based on texture zeroes studied in \cite{Borgohain}, where it is seen that the results obtained for LFV are not very satisfactory incase of two zero texture compared to that in one zero texture. Also as already mentioned in \cite{texture1}, two zero texture has been discarded from the LFV point of view. Thus, we have solely generated only one zero texture of the Yukawa coupling matrix from our model to study its significance in neutrino sector.   

	\section{Conclusion}\label{sec8}
	We have extensively studied the scotogenic model realised with the help of discrete flavor symmetries $A_{4}\otimes Z_{4}$. Our work mainly focuses on the condition required to generate texture one zero in the Yukawa coupling matrix. The various vev alignments mandatory in this aspect is been discussed in Sec.\ref{sec3}. With due change in the consideration of the vev alignments of the flavons $\chi$, $\chi^{'}$ and $\chi^{''}$, we are able to construct three different structures of Yukawa coupling matrix with a zero element in it. Since, broken $\mu-\tau$ symmetry is a crucial requirement, two structures of Yukawa coupling matrix (i.e. Case I and Case II) are forbidden and only Case III is allowed. The neutrino oscillation parameters $\theta_{12}$ and $\theta_{13}$ are also in the 3$\sigma$ global fit credible region(CR) for the allowed structure of Yukawa coupling matrix. Additionally, we take some particular range of free parameters such as $M_{1}= 10^{4}-10^{5}$ GeV, $M_{2}= 10^{6}-10^{7}$ GeV, $M_{3}= 5\times 10^{7}-10^{8}$ GeV, $m_{\eta^{0}_{R}}= 450-750$ GeV, $m_{l}= 10^{-13}-10^{-11}$ eV and $\lambda_{5}=10^{-3}- 1$ , and proceed with the calculation of various phenomena for the allowed Yukawa coupling matrix. In order to make the model feasible, we have studied the neutrino phenomenology like $0\nu\beta\beta$, lepton flavor violation and also have added a tinch of cosmology via BAU. The one zero texture matrix in eq.\ref{Y22} is evaluated in our work. % Case II tends to satisfy the KamLAND-Zen limit and Planck limit for $m_{\beta\beta}$ and BAU respectively as depicted in Fig.\ref{eff13} and Fig.\ref{bau1}.
	   We see that the allowed Case III satisfies the KamLAND-Zen limit and Planck limit for $m_{\beta\beta}$ and BAU respectively from Fig.\ref{eff22} and Fig.\ref{bau22}. Thus, from the extensive analysis we have carried out, we can consider Case III to abide by the experimental constraints alongwith a naturally broken $\mu-\tau$ symmetry. Furthermore, for the validity of the model w.r.t dark matter phenomenology, we have assumed the dark matter(lightest of the inert doublet scalar) mass $M_{DM}$ in the range 450-750 GeV. As we have considered two distinct values of the mass splittings between the inert scalars, we can draw conclusion that for the lower value of mass splitting, i.e. $\Delta M_{\eta^{\pm}}$ = $\Delta M_{\eta^{0}_{I}} = 0.9$ GeV, a wider range of allowed DM mass is obatined. The consistency of this result is shown for a benchmark value of DM-Higgs coupling $\lambda_{L}= 0.00005$ as well as for quite a broad space of  $\lambda_{L}$ as can be seen in Fig.\ref{DM} and Fig.\ref{DM1}. Thus, as a whole we can contemplate this discrete flavor realisation of the scotogenic model to be sound in explaining various beyond standard model phenomenologies and also plays a crucial role in distinguishing between the most desirable structure of the Yukawa coupling matrices.
	\section{Acknowledgement}
 The research work of MKD is supported by the Department of Science and Technology, Government of India, under the project grant EMR/2017/001436. 
	\bibliographystyle{utphys}
	\bibliography{ref1}
	
\end{document}